%% file: ANA-HIGP-2024-11-PAPER.tex
\documentclass[cernpreprint, atlasdraft=false, USenglish, texmf, orcidlogo]{atlasdoc}
\usepackage{atlaspackage}
\usepackage{atlasbiblatex}

\usepackage{atlasphysics}
\graphicspath{{logos/}{figures/}}

\usepackage{ANA-HIGP-2024-11-PAPER-defs}
\usepackage{multirow}

\AtlasTitle{Evidence for the dimuon decay of the Higgs boson in $p\!p$ collisions with the ATLAS detector}

\AtlasAbstract{%
A search for the dimuon decay of the Higgs boson is presented based on $pp$ collision data recorded by ATLAS during Run~3 of the Large Hadron Collider, corresponding to an integrated luminosity of $165~\mathrm{fb}^{-1}$ at $\sqrt{s} = 13.6~\,\text{Te\kern -0.1em V}$. To enhance the sensitivity, the results are combined with those from Run~2. An excess of events over the background is observed  with a significance of $3.4 \sigma$ ($2.5 \sigma$ expected). The best-fit signal strength is $\mu = 1.4\pm0.4$. This result provides evidence for the $H \to \mu\mu$ decay with ATLAS data and offers a direct probe of the Higgs-boson Yukawa coupling to second-generation fermions.
}

\AtlasRefCode{HIGP-2024-11}

\PreprintIdNumber{CERN-EP-2025-149}

\AtlasJournalRef{Phys. Rev. Lett. 135 (2025) 231802}
\AtlasDOI{10.1103/gzdh-p159}

\hypersetup{pdftitle={ATLAS document},pdfauthor={The ATLAS Collaboration}}

\begin{document}

\maketitle

%

%
%

%


A wealth of information about the nature of the Higgs boson, $H$, discovered in 2012 by the ATLAS and CMS Collaborations~\cite{HIGG-2012-27,CMS-HIG-12-028}, has been reaped by leveraging the Run-1 and Run-2 proton--proton ($pp$) collision datasets collected at the Large Hadron Collider (LHC)~\cite{HIGG-2021-23,HIGG-2023-11-adhoc,CMS-HIG-22-001}. With these data, the Yukawa interactions of the Higgs boson to third-generation charged fermions have been firmly established. However, its Yukawa coupling to second-generation fermions has yet to be conclusively determined. Notably, the Higgs-boson interactions are the only ones in the Standard Model (SM) that distinguish between the fermion generations.
The decay of the Higgs boson to two oppositely charged muons, \hmm, provides an excellent opportunity for a first observation of this coupling to a second-generation fermion. The SM \hmm\ branching ratio is \hmmbr~\cite{Dittmaier:2012vm} assuming a Higgs-boson mass of $m_H=125.09$~\GeV~\cite{HIGG-2014-14}.

The most recent ATLAS search for the \hmm\ decay employed the full Run-2 dataset~\cite{HIGG-2019-14}. Assuming $m_H = 125.09~\GeV$, the observed (expected) significance over the background-only hypothesis was 2.0~$\sigma$ (1.7~$\sigma$). The comparable result by CMS~\cite{CMS-HIG-19-006}, assuming $m_H = 125.38~\GeV$,  had an observed (expected) significance of 3.0~$\sigma$ (2.5~$\sigma$).

A new ATLAS search for \hmm is presented targeting the gluon-fusion (\ggF), vector-boson-fusion (VBF), associated vector-boson ($VH$), and associated top-quark-pair (\ttH) production modes. This analysis uses Run-3 $pp$ collision data with an integrated luminosity of $\lumirthree$, collected in $2022-2024$ at a center-of-mass energy \sqsrthree. The result is combined with the ATLAS Run-2 $13\TeV$ result~\cite{HIGG-2019-14}, with an integrated luminosity of $\lumirtwo$.
Apart from more than doubling the statistics of the previous result, several enhancements have been made to make the analysis more robust and performant, and include a five-billion event next-to-leading-order (NLO) Drell--Yan fully simulated sample, an improvement of the mass resolution through a \hmm decay vertex fit, the inclusion of a 2-lepton $VH$ category, the use of fully hadronic \ttH decays, and the recategorization of \ttH using a distance-correlation neural network~\cite{Kasieczka_2020}.


%

The ATLAS experiment~\cite{PERF-2007-01,GENR-2019-02} at the LHC is a multipurpose particle detector
with a forward--backward symmetric cylindrical geometry and a near \(4\pi\) coverage in
solid angle~\cite{atlasfootnote}.
It consists of an inner tracking detector surrounded by a thin superconducting solenoid, electromagnetic and hadronic calorimeters, and a muon spectrometer based on three large superconducting air-core toroidal magnets.
The data used in this analysis~\cite{SOFT-2022-02} were collected with triggers that require at least one high-transverse-momentum muon: $\pt > 24~\GeV$ and isolated, or $\pt > 50~\GeV$ without isolation requirement~\cite{TRIG-2016-01,TRIG-2022-01}. The trigger efficiency for the sum of all \hmm signal processes is about 91\% relative to the common event selection described in Table~\ref{tab:evtsel}.


%

Simulated Monte Carlo (MC) samples are used to optimize the event selection,
model the signal processes, and develop analytic functions to describe the dimuon invariant-mass distributions, \mmm, for background estimation.
Both the signal and a comprehensive set of background processes are produced using the full ATLAS detector
simulation~\cite{SOFT-2010-01}, based on \GEANT~\cite{Agostinelli:2002hh},
and are corrected to match the number of interactions per bunch crossing (in-time pileup) observed in data.
Further details are in the End Matter.


%


The physics objects necessary to reconstruct the searched-for final states consist of muons arising from the \hmm decay as well as other particles or objects produced in association with the Higgs-boson candidate, such as other muons, electrons, photons, jets, and neutrinos characterized by missing transverse momentum, \ptmiss.
The main selection requirements are summarized in \Tab{\ref{tab:evtsel}}.

\begin{table}[tb]
\begin{center}
\caption{Event selection criteria and  basic requirements on leptons, $b$-tagged jets, and \etmiss for the categories targeting Higgs-boson production processes.}
\label{tab:evtsel}
\small
\begin{tabular}{lc}
\toprule
& Selection \\
\midrule
& Primary vertex\\
Common preselection & Two opposite-charge muons\\
& Muons: $|\eta| < 2.5$, $\pT^\mathrm{lead} > 27\GeV$, $\pT^\mathrm{sublead} > 15\GeV$\\
\midrule
Fit region & $\mmm = 110- 160~\GeV$\\
\midrule
\multirow{2}{*}{Jets}            & $\pT > 25\GeV$ and $|\eta|<2.4$\\
& or with $\pT > 30\GeV$ and $2.4 < |\eta|< 4.5$ \\
\midrule
\multirow{3}{*}{$b$-tagged jets} & $\pT > 25\GeV$ and $|\eta|<2.4$\\
& or with $\pT > 30\GeV$ and $2.4 < |\eta|< 2.5$ \\
&  Tagging efficiency working point of 85\% \\
\midrule
\ttH categories & At least one $b$-jet \\
$VH$ 4-lepton category & Exactly two additional $e$ or $\mu$ with $\pT > 8\GeV, 5\GeV$ ($\mu$) / $7\GeV$ ($e$), no $b$-jets \\
$VH$ 3-lepton categories & Exactly one additional $e$ or $\mu$ with $\pT > 15\GeV$, no $b$-jets \\
$VH$ 2-lepton categories & No additional lepton, no $b$-jets, $\etmiss > 120~\GeV$\\
VBF and \ggF categories & No additional lepton, no $b$-jets, $\etmiss < 120~\GeV$\ \\
\bottomrule
\end{tabular}
\end{center}
\end{table}

Muon candidates must satisfy loose identification criteria~\cite{MUON-2018-03}, and have $\pT>5\GeV$ and pseudorapidity $|\eta|<2.5$.
The muon inner-detector track must be matched to the primary vertex by having a longitudinal impact parameter  $|z_0 \sin(\theta)|<0.5$~mm, where $\theta$ is the track polar angle, and a significance of the transverse impact parameter, $d_0$, calculated relative to the measured beam-line position, of  $|d_0|/\sigma(d_0) < 3$, where $\sigma(d_0)$ is the uncertainty in $d_0$.
Particle-flow-based isolation criteria~\cite{MUON-2018-03} are applied to suppress muons from hadron decays.
There are known correlations between \mmm and the muon-track orientation~\cite{CMS-HIG-19-006}.
By fitting the Higgs-boson decay vertex, a more precise value for $d_0$ can be determined,
yielding a better estimate of \mmm. This decay vertex is fitted with a least-squares method, that minimizes a set of estimators related to the muon tracks (track $\theta$ and $\phi$, and the charge-to-momentum ratio, $q/p$) and vertex parameters ($x,y,z$ position) with respect to the measured values, taking into account their covariances
and a constraint to keep the fitted decay vertex within the beamspot envelope.
This procedure is applied to  categories requiring exactly two muons, resulting in an improvement of the \mmm resolution of nearly 2\%.
This methodology was developed with signal \ggF/VBF and background Drell--Yan MC samples, and validated using \Zboson-boson resonance-peak 2018 data.

Muons from the \hmm decay can lose a fraction of their energy by QED final-state radiation (FSR), impacting the \mmm shape and peak position.
Photon candidates~\cite{EGAM-2021-01,EGAM-2021-02} are included in the \mmm calculation if they are found in categories requiring exactly two muons and
meet a minimum-energy requirement dependent on their closeness to the muon. An FSR photon is found for 5\% of the events.

Electrons must satisfy medium likelihood-based identification criteria~\cite{EGAM-2021-01,EGAM-2021-02}, and have $\pT>7\GeV$ and $|\eta| <2.47$.
They must be isolated from additional near-by inner-detector tracks and energy deposits in the calorimeters~\cite{EGAM-2019-01}.
Electrons must satisfy $|z_0 \sin(\theta)|<0.5$~mm and $|d_0|/\sigma(d_0) < 5$.

Jets are reconstructed from particle-flow objects~\cite{PERF-2015-09} using the anti-$k_t$ algorithm~\cite{Cacciari:2008gp,Fastjet}
with radius $R=0.4$.
To suppress contributions from pileup, separate jet-vertex-tagging~\cite{PERF-2014-03,ATL-PHYS-PUB-2019-026} requirements are applied to central and forward jets.
Jets containing $b$-hadrons are tagged  using a transformer-based neural-network discriminant GN2 tagger~\cite{FTAG-2023-05}.
The \etmiss of the event is defined as the magnitude of the negative vectorial sum of the transverse momenta of the selected and calibrated physics objects and other inner-detector tracks from the primary vertex not associated with any of these objects~\cite{JETM-2020-03}.

Events with candidate \hmm decays are selected if they contain at least two opposite-charge muons.
The signal-plus-background fits are performed in the mass range $\mmm = 110-160~\GeV$. In the mass window $\mmm= 120-130~\GeV$, which contains about $85\%$ of the signal, approximately 1000 \hmm events with the $13.6 \TeV$ data are expected, corresponding to a total efficiency times acceptance of $47\%$ with respect to all \hmm produced in the \ttH, $VH$, VBF, and \ggF processes. The leading sources of loss arise from the geometrical acceptance of the detector, the trigger, and requiring that there be at least two identified muons in the event.


%

Events satisfying the criteria of \Tab{\ref{tab:evtsel}} are classified into 23 mutually exclusive categories (see \Tab{\ref{tab:results}}) constructed to exploit the topological and kinematic differences between the background processes and the
Higgs-boson production modes under consideration: \ttH, $VH$, VBF and \ggF.
For most categories, boosted decision tree (BDT) discriminants are built using the XGBoost package~\cite{xgboost}, though the Toolkit for Multivariate Analysis (TMVA)~\cite{TMVA} is used for the 4-lepton $VH$ category and a neural network is trained using Keras~\cite{chollet2015keras} and TensorFlow~\cite{abadi2016tensorflowlargescalemachinelearning} in the  \ttH categories. The training is performed using variables sensitive to the kinematics of the final-state particles and their relations to other objects. The resulting discriminants are used to define category boundaries optimized to maximize the total number-counting significance.
To minimize biases in the categorization process, all trainings are performed using $k$-fold cross-validation, where $k=4$ different partitions are used in turn for training, validation, and testing.

\begin{table}[tbp]
\centering
\caption{
For the $\mmm = 120-130\GeV$ window, the number of observed events and the fitted number of signal ($S$) and background ($B$) events from
the combined fit to \lumirthree of Run-3 data with a signal strength of $\mu = 1.6 \pm 0.6$.
The signal uncertainty reflects the uncertainty in $\mu$, while the background uncertainty
is given by the sum in quadrature of the fit statistical uncertainty and the spurious-signal uncertainty.
Also shown are the number of signal events divided by the square root of the number of background events ($S/\sqrt{B}$)
and signal-to-background ratio ($S/B$).
}
\label{tab:results}
\small
\begin{tabular}{ l  r r@{\hskip.05cm}c@{\hskip.05cm}l r@{\hskip.05cm}c@{\hskip.05cm}l c c}
\toprule
Category             & Data         & \multicolumn{3}{c}{$S$}       &              & {$B$} &              & $S/\sqrt{B}$ & $S/B$ [\%] \\
\midrule
\ttH-High            & 12           & $1.9    $ & $\pm$ & $0.7    $ & $15.5      $ & $\pm$ & $2.3       $ & $ 0.49  $    & $ 12.5  $  \\
\ttH-Medium          & 117          & $3.9    $ & $\pm$ & $1.4    $ & $115       $ & $\pm$ & $7         $ & $ 0.36  $    & $ 3.4   $  \\
VH4L                 & 11           & $0.78   $ & $\pm$ & $0.29   $ & $12.2      $ & $\pm$ & $1.8       $ & $ 0.22  $    & $ 6.4   $  \\
VH3L-High            & 25           & $1.4    $ & $\pm$ & $0.5    $ & $17.4      $ & $\pm$ & $2.9       $ & $ 0.33  $    & $ 8.0   $  \\
VH3L-Medium          & 143          & $3.7    $ & $\pm$ & $1.4    $ & $136       $ & $\pm$ & $10        $ & $ 0.31  $    & $ 2.7   $  \\
VH2L-High            & 19           & $1.0    $ & $\pm$ & $0.4    $ & $18.3      $ & $\pm$ & $2.9       $ & $ 0.23  $    & $ 5.3   $  \\
VH2L-Medium          & 30           & $0.38   $ & $\pm$ & $0.14   $ & $31.7      $ & $\pm$ & $3.1       $ & $ 0.07  $    & $ 1.2   $  \\
VBF High             & 9            & $4.3    $ & $\pm$ & $1.6    $ & $10.5      $ & $\pm$ & $1.8       $ & $ 1.34  $    & $ 41.2  $  \\
VBF Medium           & 28           & $5.3    $ & $\pm$ & $2.0    $ & $25.8      $ & $\pm$ & $2.7       $ & $ 1.04  $    & $ 20.5  $  \\
VBF Low              & 69           & $7.2    $ & $\pm$ & $2.7    $ & $62        $ & $\pm$ & $4         $ & $ 0.91  $    & $ 11.6  $  \\
VBF Very Low         & 217          & $11     $ & $\pm$ & $4      $ & $225       $ & $\pm$ & $8         $ & $ 0.76  $    & $ 5.1   $  \\
2-jet High           & 1\,399       & $31     $ & $\pm$ & $12     $ & $1\,367    $ & $\pm$ & $25        $ & $ 0.84  $    & $ 2.3   $  \\
2-jet Medium         & 5\,657       & $69     $ & $\pm$ & $26     $ & $5\,560    $ & $\pm$ & $50        $ & $ 0.92  $    & $ 1.2   $  \\
2-jet Low            & 17\,684      & $110    $ & $\pm$ & $40     $ & $17\,300   $ & $\pm$ & $70        $ & $ 0.87  $    & $ 0.7   $  \\
2-jet Very Low       & 35\,147      & $110    $ & $\pm$ & $40     $ & $35\,160   $ & $\pm$ & $140       $ & $ 0.59  $    & $ 0.3   $  \\
1-jet High           & 708          & $17     $ & $\pm$ & $6      $ & $710       $ & $\pm$ & $16        $ & $ 0.65  $    & $ 2.4   $  \\
1-jet Medium         & 7\,166       & $80     $ & $\pm$ & $30     $ & $7\,140    $ & $\pm$ & $70        $ & $ 0.95  $    & $ 1.1   $  \\
1-jet Low            & 31\,761      & $180    $ & $\pm$ & $70     $ & $31\,510   $ & $\pm$ & $120       $ & $ 1.00  $    & $ 0.6   $  \\
1-jet Very Low       & 73\,578      & $200    $ & $\pm$ & $80     $ & $73\,330   $ & $\pm$ & $200       $ & $ 0.75  $    & $ 0.3   $  \\
0-jet High           & 19\,445      & $120    $ & $\pm$ & $50     $ & $19\,260   $ & $\pm$ & $90        $ & $ 0.89  $    & $ 0.6   $  \\
0-jet Medium         & 50\,742      & $200    $ & $\pm$ & $80     $ & $50\,830   $ & $\pm$ & $190       $ & $ 0.90  $    & $ 0.4   $  \\
0-jet Low            & 94\,032      & $240    $ & $\pm$ & $90     $ & $93\,770   $ & $\pm$ & $210       $ & $ 0.78  $    & $ 0.3   $  \\
0-jet Very Low       & 136\,762     & $170    $ & $\pm$ & $60     $ & $136\,510  $ & $\pm$ & $290       $ & $ 0.47  $    & $ 0.1   $  \\
\bottomrule
\end{tabular}
\end{table}

For each training, the targeted simulated Higgs-boson production process is considered as the signal, whereas other simulated SM processes relevant to the event-categorization topology, including other signal production modes, are used as the background.
Overall, the Drell--Yan process comprises the dominant background. Other background processes such as diboson, $t\bar{t}$, and single-top production, and rarer processes such as \ttV or electroweak production of \Zboson bosons in association with jets play significant roles in specific categories.
The  background MC is scaled so that the number of events in the sidebands ($\mmm = 110-120~\GeV$ and $\mmm = 130-160~\GeV$) matches the number of events in the data sidebands. The background yield is then taken as the number of reweighted background MC events in the window  $\mmm=120-130~\GeV$ where the expected signal-to-background ratio is computed.
The category selections are performed in the same order as presented in this Letter: \ttH, $VH$, VBF, and \ggF.

The \ttH category targets both hadronic and leptonic final states arising from the decay of the \Wboson boson produced in the top-quark decay.
The two highest-\pt muons with opposite charge are chosen as the Higgs-boson decay candidate. This procedure correctly selects the Higgs-boson muon pair in 91\% of the cases.
Thirty variables are used as input to train a neural network that  uses a custom loss function consisting of a term that calculates the binary cross-entropy between the neural network predictions and assigns each event as signal or background, and a term that measures the distance correlation between the neural-network predictions and the dimuon invariant mass of background events~\cite{Kasieczka_2020}. This latter term suppresses the shaping of the mass spectrum when defining categories based on the neural-network score, by penalizing any correlation between the scores and the mass.
The resulting output-score distribution is used to define two regions with the highest significance: \ttH-High (99\% purity) and \ttH-Medium (96\% purity), where the quoted purity is with respect to all Higgs-boson signals.

Events not selected in the \ttH category are considered for the $VH$-enriched categories targeting events where the Higgs boson is produced in association with a leptonically decaying vector boson, \wln or \zll , where $\ell=e,\mu$, and $Z \rightarrow \nu \nu$ decays.
BDTs are trained separately to select five categories according to the lepton multiplicity: 4, 3, or 2.
Events entering the 4-lepton category must contain exactly four well-separated and isolated leptons, either four muons or two muons and two electrons.
The opposite-charge same-flavor dilepton pair with mass closest to the \Zboson boson is labeled as the \Zboson-boson candidate, and the remaining opposite-charge pair is labeled as the Higgs-boson candidate, if these leptons are muons.
The accuracy of the Higgs-boson muon-pair assignment is estimated to be nearly 99\% within $\mmm = 120-130 \GeV$.
Fifteen variables are used as input for the BDT training.
In the 3-lepton channel for the case where the additional lepton is a muon, the assignment of leptons to either Higgs-boson or \Wboson-boson decay candidates is based on a $\chi^2$  minimization criterion that takes into account the lepton charge, the difference between the reconstructed and expected Higgs-boson mass and \Wboson-boson transverse mass (built from the lepton $\vec{p}_\text{T}$ and the event \ptmiss), and their expected experimental resolutions.
The correct pairing is obtained in 96\% of the cases.
Isolation requirements are tightened on the lepton assigned to the \Wboson-boson candidate.
Twelve variables are used as input for the BDT.
The 2-lepton channel must contain exactly two well-isolated muons and uses
9 variables as input for the BDT training.
In the end, five $VH$ categories are defined by applying requirements on the BDT scores: one for the 4-lepton channel (VH4L) and two each for the 3- and 2-lepton channels (VH3L-High, VH3L-Medium, VH2L-High, and VH2L-Medium).
The expected \hmm signal purity for the $VH$ production process ranges from 99\% in VH4L to  63\% for VH2L-Medium.

The events not selected in the \ttH or $VH$ categories  are further classified according to the number of reconstructed jets in the event: 0-jet, 1-jet and $\ge 2$-jet (the latter referred to as `2-jet').
The BDT training applied to each jet-multiplicity category uses 17 variables.
A VBF classifier is built from 2-jet events and four categories with the highest BDT scores are defined: VBF High, Medium, Low, and Very Low, with purity ranging from 96\% (High) to 76\% (Very Low).
The remaining events are  considered for classification with BDTs split by jet multiplicity, and
selected on the BDT scores, to yield 12 mutually exclusive categories: $N$-jet High, Medium, Low and Very Low, where $N=0,1,2$. The expected purity for \ggF production ranges from  >99\% to 67\%.


%

Signal extraction is performed with a binned maximum-likelihood fit to \mmm in the $110-160~\GeV$ interval.
Analytical models are used in the fit to describe both signal and background, following the methodology used in the corresponding ATLAS Run-2 search~\cite{HIGG-2019-14}.

The Higgs-boson signal is modeled using a double-sided Crystal Ball~\cite{Oreglia:1980cs, HIGG-2014-04} probability density function (pdf), fitted to the total expected signal in each category.
The width of the Gaussian core, parametrizing the \mmm resolution, changes with kinematics from 2.8~$\GeV$ in the 0-jet categories to 3.2~$\GeV$ in the $VH$ and \ttH categories.

Systematic uncertainties in the signal modeling arise from both theory and experimental sources,
and are incorporated into the signal-extraction fit via nuisance parameters
that affect the signal yields and mass distributions in each category.
Theory uncertainties primarily affect the expected number of signal events.
Uncertainties in the predicted SM production cross-sections and branching ratios follow the recommendations of \Refns{\cite{deFlorian:2016spz}}{\cite{ATL-PHYS-PUB-2023-031}}.
Uncertainties due to missing higher orders in Quantum Chromodynamic (QCD) calculations are estimated based on the Stage 1.2 simplified template cross-sections (STXS) scheme~\cite{deFlorian:2016spz,badger2016leshouches2015physics,berger2019simplifiedtemplatecrosssections}.
For all signal processes, parton distribution function (PDF) uncertainties are evaluated by using eigenvector variations of the NNPDF2.3LO set~\cite{Ball:2012cx}.
For the \ggF\ and VBF processes, modeling uncertainties related to the underlying event
and parton showers are assessed by comparing simulations from \PYTHIA{8}~\cite{Sjostrand:2014zea} and \HERWIG{7}~\cite{Bahr:2008pv,Bellm:2015jjp}.

Experimental systematic uncertainties arise from muon-related efficiency uncertainties associated with reconstruction, identification, trigger, isolation, and impact-parameter requirements~\cite{MUON-2018-03, TRIG-2018-01} and uncertainties in the muon momentum scale and resolution~\cite{MUON-2022-01}, both of which significantly affect the shape of the signal mass distribution.

Additional experimental uncertainties include those related to \etmiss\ ~\cite{JETM-2020-03},
the $b$-tagging efficiency~\cite{FTAG-2018-01,FTAG-2019-02,FTAG-2020-08}, the modeling of pile-up interactions~\cite{ATL-PHYS-PUB-2016-017},
the measurement of electron reconstruction and identification efficiencies~\cite{EGAM-2021-01,EGAM-2021-02},
the calibration of photon energy scale~\cite{EGAM-2021-01,EGAM-2021-02}, and the data luminosity estimation~\cite{lumifootnote,ATL-DAPR-PUB-2024-001}.
Furthermore, uncertainties in the jet-reconstruction efficiency, energy scale, and resolution~\cite{PERF-2016-04} are included.
To account for observed mismodeling of jets at large pseudorapidity ($|\eta| > 3.2$),
an additional systematic uncertainty is introduced, defined as five times the difference between data and Monte Carlo event yields
resulting from the application of the energy shift measured in the forward calorimeters.

The $\mmm$ background-modeling strategy and the extraction of the main systematic uncertainty (spurious signal) follow Ref.~\cite{HIGG-2019-14}.
The main differences with respect to this reference are the use of high-statistics, fully-simulated NLO samples as the sole background template,
and an improved core function—based on an analytical leading-order (LO) Drell--Yan line shape—that includes a more accurate treatment
of experimental dimuon mass-resolution effects and accounts for parton luminosity contributions from $b$- and $c$-quark initial states.
Typical background parametrizations include exponentials of polynomials and power-law functions, with degrees of freedom ranging from 1 to 4, and the spurious signal is generally within the range of 5–40\% of the expected signal statistical uncertainty.

The search is dominated by statistical uncertainties,
with systematic uncertainties accounting for approximately 5\% of the total.
The leading systematic contribution arises from background modeling,
followed by theory uncertainties, muon-related effects, and the knowledge of the integrated luminosity.


%

\begin{figure}[tbp]
\centering
\includegraphics[width=0.7\textwidth]{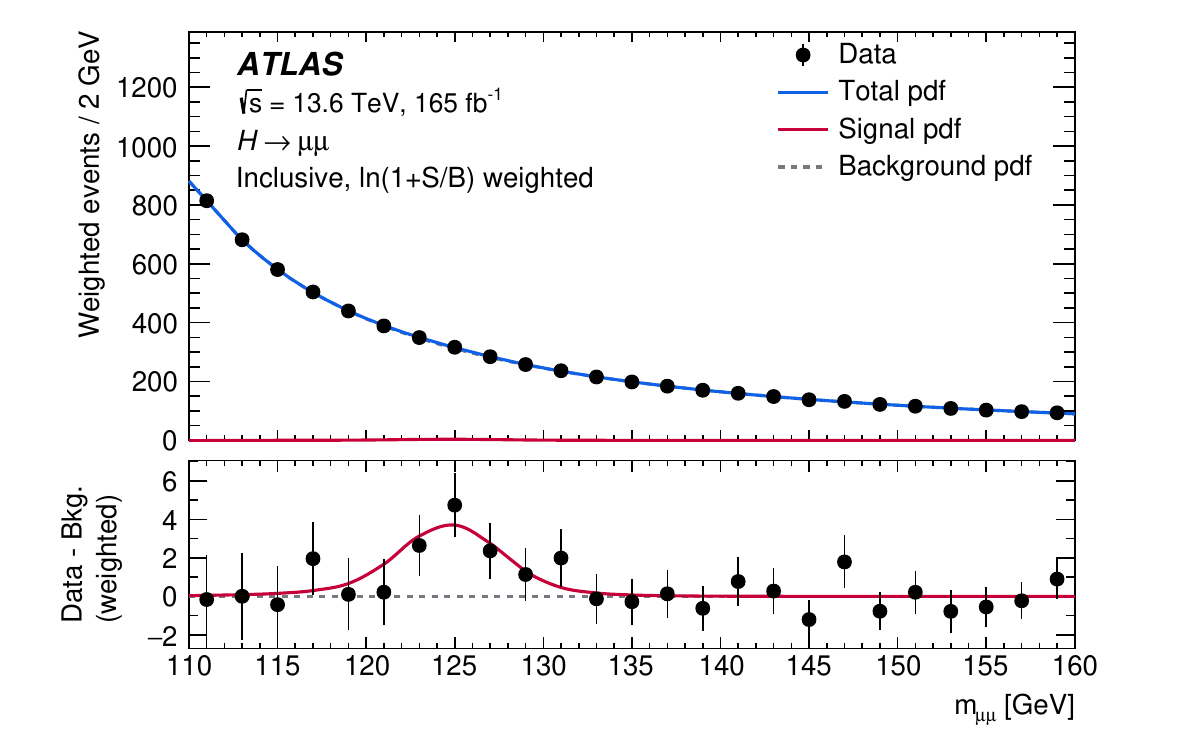}
\caption{
Observed dimuon invariant mass spectrum combining all of the Run-3 categories.
The events and probability density functions (pdf) are weighted by
$\ln(1 + S/B)$, where $S$ and $B$ are the signal and background yields, respectively,
evaluated in the $\mmm = 120-130\,\GeV$ window.
These yields are extracted from the fit performed in the mass range $\mmm = 110-160~\GeV$,
providing a signal strength of $\mus = 1.6 \pm 0.6$.
The background and signal pdfs are obtained from the fit to the data.
The lower panel shows the data with the background model subtracted,
and the fitted signal pdf, normalized to the signal best-fit value.
Error bars represent the data statistical uncertainties.
}
\label{fig:inclusive}
\end{figure}

The signal-plus-background model is used to fit to the observed \mmm spectra in the 23 categories simultaneously.
Events in \Fig{\ref{fig:inclusive}} are weighted by $\ln(1+S/B)$, with $S$ and $B$ denoting the observed signal and background yields, respectively, as extracted from the fit to data
in the $\mmm=120-130\GeV$ range.
The corresponding values for $S$, $B$, and other relevant quantities for each category are summarized in \Tab{\ref{tab:results}}.

\begin{figure}[tbp]
\centering
\includegraphics[width=0.7\textwidth]{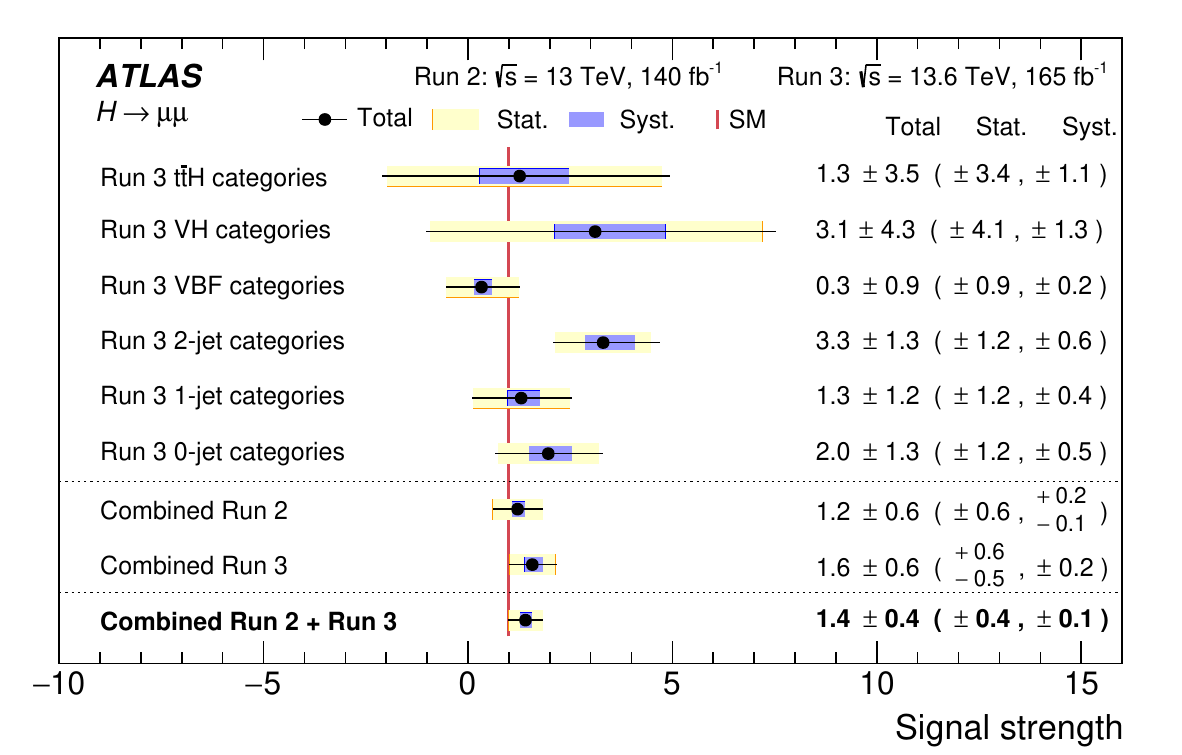}
\caption{
Best-fit values of the signal-strength parameters for the analysis of the Run-3 data and the full Run-2 dataset.
Results are first shown for \rthree by grouping the signal strength parameters into six major categories (\ttH, $VH$, VBF, 2-jet, 1-jet, and 0-jet), where the systematic uncertainties reported in the legend are symmetrized.
In the lower part of the plot, the full Run-2 and Run-3 results are given, as well as their combination.}
\label{fig:mu}
\end{figure}

The best-fit value of the signal-strength parameter, defined as the ratio of the observed signal yield to the SM expectation,
is $\mus = 1.6 \pm 0.6$, and corresponds to an observed (expected) significance of $2.8$ ($1.8$) standard deviations, $\sigma$,
with respect to the background-only hypothesis.
The measured signal strengths in each group of categories, together with the combined Run-3 result, are shown in \Fig{\ref{fig:mu}}.
The compatibility of the measured signal strengths between the 23 categories is tested
by repeating the fit after allowing each category to have its own signal strength parameter.
The probability of compatibility is at the 68\% level.
Using the same methodology, the probability of compatibility between the signal strengths of the six groups
of categories shown in \Fig{\ref{fig:mu}} is 54\%.

The results presented in this Letter are combined with the analysis of the full ATLAS Run-2 dataset~\cite{HIGG-2019-14}.
Experimental uncertainties are taken as uncorrelated between the two datasets,
while theory sources are treated as fully correlated, except for those related to PDFs.
An excess of events over background is observed with an observed (expected) significance of 3.4 (2.5)~$\sigma$.
The measured signal strength in the combined fit is $\mus = 1.4 \pm 0.4$,
and the individual results for \rtwo and \rthree are compatible within 68\%.
The measured signal strengths for the Run-2 and Run-3 datasets, as well as their combination, are shown in \Fig{\ref{fig:mu}}.
Assuming the SM values for the Higgs-boson production cross-sections, this result corresponds to a branching fraction of $\mathrm{B}(\hmm) = (3.0 \pm 0.9) \times 10^{-4}$.

%

In conclusion, the ATLAS experiment sees evidence for the rare decay of the Higgs boson into two muons,  and hence for the  Higgs-boson Yukawa coupling to second-generation fermions,
with an observed (expected) significance over the background-only hypothesis of 3.4 (2.5)~$\sigma$.
The result is obtained by analysing \lumirthree of $pp$ data collected
at $\sqrt{s} = 13.6~\TeV$ and combining with the results from \lumirtwo of data collected at $\sqrt{s} = 13~\TeV$.
This result is compatible with the SM expectation.

\clearpage


%

%
%
%
%
%

%
\section*{Acknowledgements}

%
%

%
%

We thank CERN for the very successful operation of the LHC and its injectors, as well as the support staff at
CERN and at our institutions worldwide without whom ATLAS could not be operated efficiently.

The crucial computing support from all WLCG partners is acknowledged gratefully, in particular from CERN, the ATLAS Tier-1 facilities at TRIUMF/SFU (Canada), NDGF (Denmark, Norway, Sweden), CC-IN2P3 (France), KIT/GridKA (Germany), INFN-CNAF (Italy), NL-T1 (Netherlands), PIC (Spain), RAL (UK) and BNL (USA), the Tier-2 facilities worldwide and large non-WLCG resource providers. Major contributors of computing resources are listed in Ref.~\cite{ATL-SOFT-PUB-2025-001}.

We gratefully acknowledge the support of ANPCyT, Argentina; YerPhI, Armenia; ARC, Australia; BMWFW and FWF, Austria; ANAS, Azerbaijan; CNPq and FAPESP, Brazil; NSERC, NRC and CFI, Canada; CERN; ANID, Chile; CAS, MOST and NSFC, China; Minciencias, Colombia; MEYS CR, Czech Republic; DNRF and DNSRC, Denmark; IN2P3-CNRS and CEA-DRF/IRFU, France; SRNSFG, Georgia; BMFTR, HGF and MPG, Germany; GSRI, Greece; RGC and Hong Kong SAR, China; ICHEP and Academy of Sciences and Humanities, Israel; INFN, Italy; MEXT and JSPS, Japan; CNRST, Morocco; NWO, Netherlands; RCN, Norway; MNiSW, Poland; FCT, Portugal; MNE/IFA, Romania; MSTDI, Serbia; MSSR, Slovakia; ARIS and MVZI, Slovenia; DSI/NRF, South Africa; MICIU/AEI, Spain; SRC and Wallenberg Foundation, Sweden; SERI, SNSF and Cantons of Bern and Geneva, Switzerland; NSTC, Taipei; TENMAK, T\"urkiye; STFC/UKRI, United Kingdom; DOE and NSF, United States of America.

Individual groups and members have received support from BCKDF, CANARIE, CRC and DRAC, Canada; CERN-CZ, FORTE and PRIMUS, Czech Republic; COST, ERC, ERDF, Horizon 2020, ICSC-NextGenerationEU and Marie Sk{\l}odowska-Curie Actions, European Union; Investissements d'Avenir Labex, Investissements d'Avenir Idex and ANR, France; DFG and AvH Foundation, Germany; Herakleitos, Thales and Aristeia programmes co-financed by EU-ESF and the Greek NSRF, Greece; BSF-NSF and MINERVA, Israel; NCN and NAWA, Poland; La Caixa Banking Foundation, CERCA Programme Generalitat de Catalunya and PROMETEO and GenT Programmes Generalitat Valenciana, Spain; G\"{o}ran Gustafssons Stiftelse, Sweden; The Royal Society and Leverhulme Trust, United Kingdom.

In addition, individual members wish to acknowledge support from CERN: European Organization for Nuclear Research (CERN DOCT); Chile: Agencia Nacional de Investigaci\'on y Desarrollo (FONDECYT 1230812, FONDECYT 1240864, Fondecyt 3240661); China: Chinese Ministry of Science and Technology (MOST-2023YFA1605700, MOST-2023YFA1609300), National Natural Science Foundation of China (NSFC - 12175119, NSFC 12275265); Czech Republic: Czech Science Foundation (GACR - 24-11373S), Ministry of Education Youth and Sports (ERC-CZ-LL2327, FORTE CZ.02.01.01/00/22\_008/0004632), PRIMUS Research Programme (PRIMUS/21/SCI/017); EU: H2020 European Research Council (ERC - 101002463); European Union: European Research Council (BARD No. 101116429, ERC - 948254, ERC 101089007), European Regional Development Fund (SMASH COFUND 101081355, SLO ERDF), Horizon 2020 Framework Programme (MUCCA - CHIST-ERA-19-XAI-00), European Union, Future Artificial Intelligence Research (FAIR-NextGenerationEU PE00000013), Italian Center for High Performance Computing, Big Data and Quantum Computing (ICSC, NextGenerationEU); France: Agence Nationale de la Recherche (ANR-21-CE31-0022, ANR-22-EDIR-0002); Germany: Baden-Württemberg Stiftung (BW Stiftung-Postdoc Eliteprogramme), Deutsche Forschungsgemeinschaft (DFG - 469666862, DFG - CR 312/5-2); China: Research Grants Council (GRF); Italy: Istituto Nazionale di Fisica Nucleare (ICSC, NextGenerationEU), Ministero dell'Università e della Ricerca (NextGenEU 153D23001490006 M4C2.1.1, NextGenEU I53D23000820006 M4C2.1.1, NextGenEU I53D23001490006 M4C2.1.1, SOE2024\_0000023); Japan: Japan Society for the Promotion of Science (JSPS KAKENHI JP22H01227, JSPS KAKENHI JP22H04944, JSPS KAKENHI JP22KK0227, JSPS KAKENHI JP24K23939, JSPS KAKENHI JP24KK0251, JSPS KAKENHI JP25H00650, JSPS KAKENHI JP25H01291, JSPS KAKENHI JP25K01023); Norway: Research Council of Norway (RCN-314472); Poland: Ministry of Science and Higher Education (IDUB AGH, POB8, D4 no 9722), Polish National Science Centre (NCN 2021/42/E/ST2/00350, NCN OPUS 2023/51/B/ST2/02507, NCN OPUS nr 2022/47/B/ST2/03059, NCN UMO-2019/34/E/ST2/00393, UMO-2022/47/O/ST2/00148, UMO-2023/49/B/ST2/04085, UMO-2023/51/B/ST2/00920, UMO-2024/53/N/ST2/00869); Portugal: Foundation for Science and Technology (FCT); Spain: Ministry of Science and Innovation (MCIN \& NextGenEU PCI2022-135018-2, MICIN \& FEDER PID2021-125273NB, RYC2019-028510-I, RYC2020-030254-I, RYC2021-031273-I, RYC2022-038164-I); Sweden: Carl Trygger Foundation (Carl Trygger Foundation CTS 22:2312), Swedish Research Council (Swedish Research Council 2023-04654, VR 2021-03651, VR 2022-03845, VR 2022-04683, VR 2023-03403, VR 2024-05451), Knut and Alice Wallenberg Foundation (KAW 2018.0458, KAW 2022.0358, KAW 2023.0366); Switzerland: Swiss National Science Foundation (SNSF - PCEFP2\_194658); United Kingdom: Royal Society (NIF-R1-231091); United States of America: U.S. Department of Energy (ECA DE-AC02-76SF00515), Neubauer Family Foundation.

%
%


\clearpage
%

\clearpage
\appendix
\part*{End Matter}
\addcontentsline{toc}{part}{Appendix}

\section{Simulated samples}
\label{app:samples}

Signal samples are generated for the primary Higgs-boson production modes,
with $\mhiggs=\samplesmh$ and $\gammahiggs=\samplesgh$~\cite{Heinemeyer:2013tqa},
using the \POWHEG~\cite{Nason:2004rx,Frixione:2007vw,Alioli:2010xd,Campbell:2012am,Hamilton:2012np,Hamilton:2012rf,Hamilton:2013fea,Hamilton:2015nsa}
program and employing the NNPDF2.3LO parton distribution function set~\cite{Ball:2012cx}.
Parton-level events are processed with \PYTHIA[8]~\cite{Sjostrand:2014zea} to perform Higgs-boson decays,
parton showering, final-state QED photon radiation, hadronization, and modeling of the underlying event,
using the A14 tune of parameters~\cite{ATL-PHYS-PUB-2014-021}.
These samples are normalized using the most up-to-date theory predictions for the SM production cross-sections,
as summarized in Ref.~\cite{deFlorian:2016spz}. The normalization also includes the \hmm branching ratio of \hmmbr,
computed using HDECAY~\cite{Djouadi:1997yw,Spira:1997dg,Djouadi:2006bz,Djouadi_2019}
and PROPHECY4F~\cite{Bredenstein:2006ha,Bredenstein:2006rh,Bredenstein:2006nk}.

The simulation of Higgs-boson production via the gluon-fusion process attains
next-to-next-to-leading-order (NNLO) QCD accuracy for inclusive observables using the NNLOPS prescription and
after reweighting the Higgs-boson rapidity spectrum~\cite{Catani:2007vq}.
The sample normalization is based on a next-to-NNLO QCD calculation
including next-to-leading-order (NLO) electroweak corrections~\cite{Aglietti:2004nj,Actis:2008ug,Actis:2008ts,Anastasiou:2008tj,Pak:2009dg,Harlander:2009bw,Harlander:2009mq,Harlander:2009my,Actis:2008ug,Anastasiou:2015ema,Anastasiou:2016cez,Dulat:2018rbf,Bonetti:2018ukf}.

Samples for Higgs-boson production via the vector-boson-fusion process and in association with a vector boson, $q\bar{q}/qg \to VH$ with $V=W$ or $Z$,
are generated using the \POWHEGBOX program~\cite{Nason:2009ai,Cullen:2011ac,Luisoni:2013cuh} at NLO accuracy, with the sole exception of
the loop-induced process $gg \to ZH$, which is generated at leading order (LO).
VBF samples are normalized to an approximate-NNLO QCD cross-section with NLO electroweak corrections~\cite{Ciccolini:2007jr,Ciccolini:2007ec,Bolzoni:2010xr},
while $VH$ samples are normalized to cross-sections calculated at NNLO in QCD with NLO electroweak corrections for $q\bar{q}/qg \to VH$ and
at NLO and next-to-leading-logarithm accuracy in QCD for $gg \to ZH$~\cite{Ciccolini:2003jy,Brein:2003wg,Brein:2011vx,Altenkamp:2012sx,Denner:2014cla,Brein:2012ne,Harlander:2014wda,Harlander:2018yio}.

Higgs-boson production in association with a top-quark pair, \ttH, is simulated at NLO accuracy in QCD and is normalized
using a cross-section calculation that is accurate to NLO in QCD with NLO
electroweak corrections~\cite{Beenakker:2002nc,Dawson:2003zu,Yu:2014cka,Frixione:2015zaa}.

Higgs-boson production in association with a bottom-quark pair, \bbH, is simulated at NLO accuracy in QCD using the four-flavor scheme.
The simulation is normalized using a cross-section computed at NNLO in the five-flavor scheme and at NLO in QCD in the four-flavor scheme,
including NLO electroweak corrections~\cite{karlberg2024adinterimrecommendationshiggs}.

Background events from the Drell--Yan $Z/\gamma^* \to\mu\mu$ process
are generated with \SHERPA[2.2.14]~\cite{Bothmann:2019yzt}.
Drell--Yan events are generated using NLO-accurate matrix elements for up to two partons,
and LO-accurate matrix elements for up to five partons calculated with the \textsc{Comix}~\cite{Gleisberg:2008fv}
and \OPENLOOPS~\cite{Cascioli:2011va,Denner:2016kdg} libraries and the NNPDF3.0 NNLO set~\cite{Ball:2014uwa}.
The events are matched to the \SHERPA parton shower~\cite{Schumann:2007mg} using the MEPS@NLO
prescription~\cite{Hoeche:2011fd,Hoeche:2012yf,Catani:2001cc,Hoeche:2009rj}.
Diboson ($WW$, $WZ$, and $ZZ$) and electroweak $Zjj$ events are produced with \SHERPA[2.2.14] and \SHERPA[2.2.16],
with a similar setup as Drell--Yan and NLO-accurate matrix elements for up to one parton and LO-accurate matrix elements for up to three partons.

The $t\bar{t}$ and single-top-quark samples are generated at NLO accuracy with
\POWHEGBOX~\cite{Frixione:2007nw,Re:2010bp} using the NNPDF3.0NLO PDF set.
They are interfaced to \PYTHIA[8] for parton showering and hadronization using
the A14 parameter set and the NNPDF2.3LO PDF set.


%
%
%
\printbibliography
\clearpage
\input{atlas_authlist}
%

%
%

%

%
%
%
%

\end{document}

%% file: atlas_authlist.tex
 
\begin{flushleft}
\hypersetup{urlcolor=black}
{\Large The ATLAS Collaboration}

\bigskip

\AtlasOrcid[0000-0002-6665-4934]{G.~Aad}$^\textrm{\scriptsize 104}$,
\AtlasOrcid[0000-0001-7616-1554]{E.~Aakvaag}$^\textrm{\scriptsize 17}$,
\AtlasOrcid[0000-0002-5888-2734]{B.~Abbott}$^\textrm{\scriptsize 123}$,
\AtlasOrcid[0000-0002-0287-5869]{S.~Abdelhameed}$^\textrm{\scriptsize 119a}$,
\AtlasOrcid[0000-0002-1002-1652]{K.~Abeling}$^\textrm{\scriptsize 55}$,
\AtlasOrcid[0000-0001-5763-2760]{N.J.~Abicht}$^\textrm{\scriptsize 49}$,
\AtlasOrcid[0000-0002-8496-9294]{S.H.~Abidi}$^\textrm{\scriptsize 30}$,
\AtlasOrcid[0009-0003-6578-220X]{M.~Aboelela}$^\textrm{\scriptsize 45}$,
\AtlasOrcid[0000-0002-9987-2292]{A.~Aboulhorma}$^\textrm{\scriptsize 36e}$,
\AtlasOrcid[0000-0001-5329-6640]{H.~Abramowicz}$^\textrm{\scriptsize 157}$
% ...
\bigskip
\\

% ...
$^{17}$Department for Physics and Technology, University of Bergen, Bergen; Norway.\\
% ...
$^{30}$Physics Department, Brookhaven National Laboratory, Upton NY; United States of America.\\
% ...
$^{36}$$^{(a)}$Facult\'e des Sciences Ain Chock, Universit\'e Hassan II de Casablanca;$^{(b)}$Facult\'{e} des Sciences, Universit\'{e} Ibn-Tofail, K\'{e}nitra;$^{(c)}$Facult\'e des Sciences Semlalia, Universit\'e Cadi Ayyad, LPHEA-Marrakech;$^{(d)}$LPMR, Facult\'e des Sciences, Universit\'e Mohamed Premier, Oujda;$^{(e)}$Facult\'e des sciences, Universit\'e Mohammed V, Rabat;$^{(f)}$Institute of Applied Physics, Mohammed VI Polytechnic University, Ben Guerir; Morocco.\\
% ...
$^{45}$Physics Department, Southern Methodist University, Dallas TX; United States of America.\\
% ...
$^{49}$Fakult\"{a}t Physik , Technische Universit{\"a}t Dortmund, Dortmund; Germany.\\
% ...
$^{55}$II. Physikalisches Institut, Georg-August-Universit\"{a}t G\"ottingen, G\"ottingen; Germany.\\
% ...
$^{104}$CPPM, Aix-Marseille Universit\'e, CNRS/IN2P3, Marseille; France.\\
% ...
$^{119}$$^{(a)}$New York University Abu Dhabi, Abu Dhabi;$^{(b)}$United Arab Emirates University, Al Ain; United Arab Emirates.\\
% ...
$^{123}$Homer L. Dodge Department of Physics and Astronomy, University of Oklahoma, Norman OK; United States of America.\\
% ...
$^{157}$Raymond and Beverly Sackler School of Physics and Astronomy, Tel Aviv University, Tel Aviv; Israel.\\
% ...
$^{a}$ Also at Affiliated with an institute formerly covered by a cooperation agreement with CERN.\\
$^{b}$ Also at An-Najah National University, Nablus; Palestine.\\
$^{c}$ Also at Borough of Manhattan Community College, City University of New York, New York NY; United States of America.\\
$^{d}$ Also at Center for Interdisciplinary Research and Innovation (CIRI-AUTH), Thessaloniki; Greece.\\
$^{e}$ Also at Centre of Physics of the Universities of Minho and Porto (CF-UM-UP); Portugal.\\
$^{f}$ Also at CERN, Geneva; Switzerland.\\
% ...

\end{flushleft}
